\documentclass[aps,prl,showpacs,floatfix,tightenlines,twocolumn,amsmath,amssymb,nobalancelastpage,reprint,superscriptaddress]{revtex4-1}
\usepackage{graphicx}
\usepackage{color}
\usepackage{ulem}
\usepackage{bm}
\usepackage{hyperref}
\usepackage{setspace}

\newcommand{\LZMO}{LiZn$_2$Mo$_3$O$_8$}
\newcommand{\ZMO}{Zn$_2$Mo$_3$O$_8$}
\newcommand{\MO}{Mo$_3$O$_{13}$}

\def\etal{\textit{et al.}}

\def\jpcm{J.\ Phys.:\ Condens.\ Matter }

\def\jpsj{J.\ Phys.\ Soc.\ Jpn.\ }

\def\jssc{J.\ Sol.\ St.\ Chem.\ }

\def\natmat{Nature Mater.\ }
\def\natphys{Nature Physics }

\def\prb{Phys.\ Rev.\ B }
\def\prl{Phys.\ Rev.\ Lett.\ }

\begin{document}
%---------------------------------------------------------------------
\title{Molecular Quantum Magnetism in \LZMO }
% ------------------
\author{M. Mourigal}
\affiliation{Institute for Quantum Matter 
             and Department of Physics and Astronomy,
             The Johns Hopkins University, Baltimore, MD 21218, USA}
% ------------------
\author{W. T. Fuhrman}
\affiliation{Institute for Quantum Matter 
             and Department of Physics and Astronomy,
             The Johns Hopkins University, Baltimore, MD 21218, USA}		
% ------------------ 
\author{J. P. Sheckelton}
\affiliation{Institute for Quantum Matter 
             and Department of Physics and Astronomy,
             The Johns Hopkins University, Baltimore, MD 21218, USA}
\affiliation{Department of Chemistry,
             The Johns Hopkins University, Baltimore, MD 21218, USA}										
% ------------------ 						
\author{A. Wartelle}
\affiliation{Institute for Quantum Matter 
             and Department of Physics and Astronomy,
             The Johns Hopkins University, Baltimore, MD 21218, USA}  
\affiliation{\'Ecole Normale Sup\'erieure de Lyon, Universit\'e de Lyon, 
						  46 All\'ee d'Italie, 69364 Lyon Cedex 07, France}
% ------------------
\author{J. A. Rodriguez-Rivera}
\affiliation{NIST Center for Neutron Research, National Institute of Standards and Technology, 
						Gaithersburg, MD 20899, USA}
\affiliation{Department of Materials Science and Engineering, University of Maryland, College 		
						 Park, MD 20742, USA}		
% ------------------																
\author{D. L. Abernathy}				
\affiliation{Quantum Condensed Matter Division, Oak Ridge National Laboratory, Oak Ridge, 
						 Tennessee 37831-6475, USA}				
% ------------------
\author{T. M. McQueen}
\affiliation{Institute for Quantum Matter 
             and Department of Physics and Astronomy,
             The Johns Hopkins University, Baltimore, MD 21218, USA}
\affiliation{Department of Chemistry,
             The Johns Hopkins University, Baltimore, MD 21218, USA}	
% ------------------
\author{C. L. Broholm}
\affiliation{Institute for Quantum Matter 
             and Department of Physics and Astronomy,
             The Johns Hopkins University, Baltimore, MD 21218, USA}
\affiliation{NIST Center for Neutron Research, National Institute of Standards and Technology, 
						 Gaithersburg, MD 20899, USA}		
\affiliation{Quantum Condensed Matter Division, Oak Ridge National Laboratory, Oak Ridge, 
						 Tennessee 37831-6475, USA}										
% ------------------    
\date{December 10, 2013}

% --------------------------------------------------------------------
\begin{abstract}
Inelastic neutron scattering for temperatures below 30 K from a powder of LiZn$_2$Mo$_3$O$_8$ demonstrates this triangular-lattice antiferromagnet hosts collective magnetic excitations from spin 1/2 Mo$_3$O$_{13}$ molecules. Apparently gapless ($\Delta<0.2$ meV) and extending at least up to 2.5 meV, the low energy magnetic scattering cross section is surprisingly broad in momentum space and  involves one third of the spins present above 100 K. The data are compatible with the presence of valence-bonds involving nearest-neighbor and next-nearest-neighbor spins forming a disordered or dynamic state.
\end{abstract}
% --------------------------------------------------------------------

% --------------------------------------------------------------------
% --------------------------------------------------------------------
\maketitle
% --------------------------------------------------------------------

Insulating spin systems have demonstrated their potential to host new states of matter emerging from electronic correlations, quantum fluctuations and entanglement~\cite{Anderson73, Lee08}. They offer a unique and controlled avenue for quantitative comparisons between quantum many-body theory and experimental observations~\cite{Stone03,Lake10,Coldea10}. Considerable theoretical efforts are now devoted to understanding the ground-state and excitations of two-dimensional (2D) antiferromagnets where spin interactions are frustrated as for the triangular and Kagome lattices~\cite{Hao09,Balents10}. Neutron scattering investigations of materials with such lattices discovered important features associated with the concept of the quantum spin-liquid~\cite{Balents10}, such as the absence of static correlations down to very low temperatures~\cite{deVries09,Fak12} and deconfined fractional spin excitations~\cite{Coldea01, Han12}. 

%% --------------------------------------------------------------------
\begin{figure}[t!]
\includegraphics[width=0.90\columnwidth]{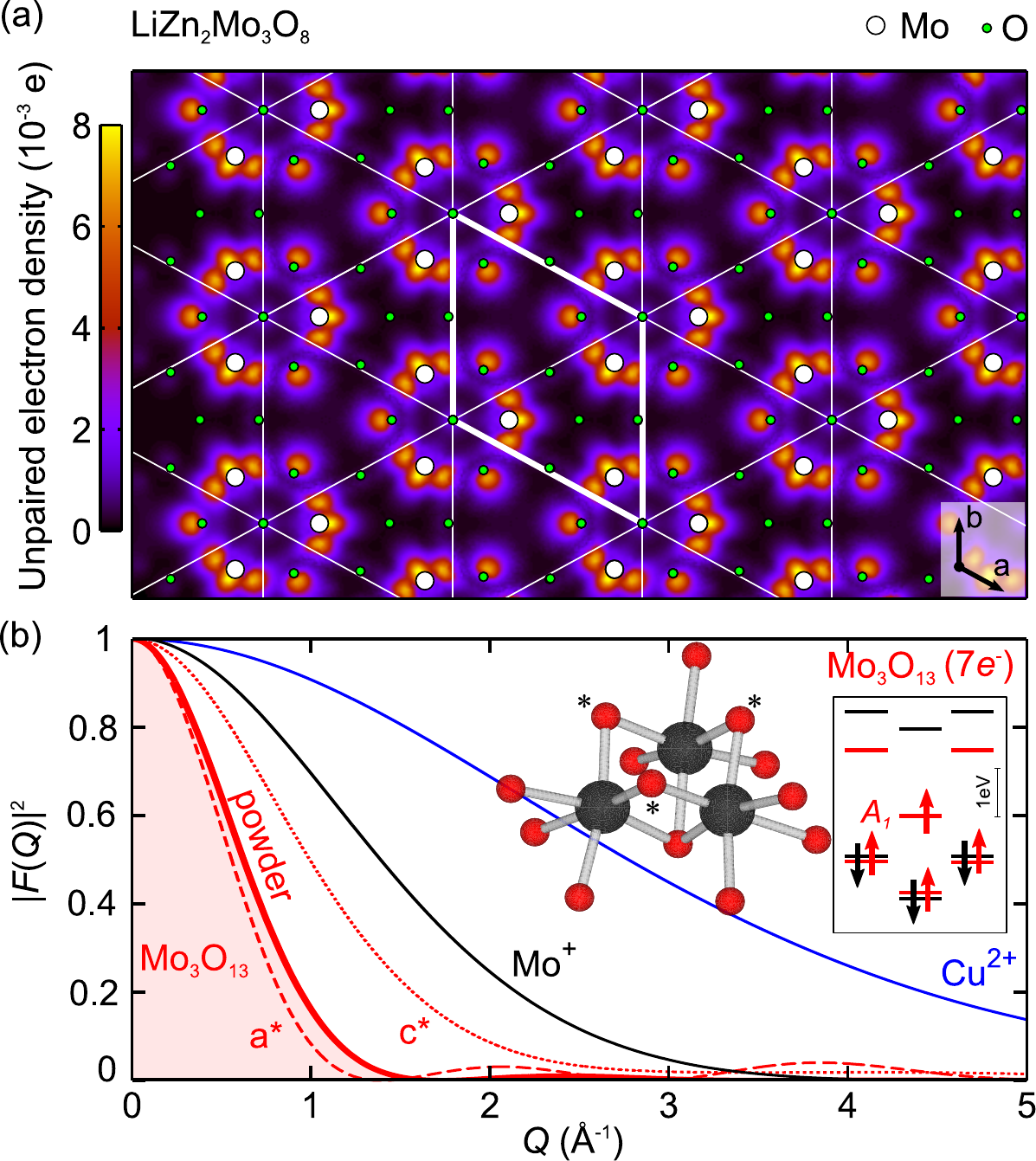}
\caption{(Color online) (a) Triangular lattice of \MO\ molecules. Thick lines highlight the unit-cell and the colormap the unpaired electronic density of \MO(7$e$) integrated along $c$. (b) Square of the spin-only form-factor of \MO(7$e$) (molecular orbital level diagram in inset) calculated along $a^*$ (dashed line), $c^*$ (dotted line), powder averaged (solid line) and compared to that of Cu$^{2+}$ (blue thin line) and Mo$^+$ (black thin line). \vspace{-0.5cm}}
\label{fig1}
\end{figure}
%% --------------------------------------------------------------------

To make quantitative comparisons with theory, the sensitivity to defects and site mixing inherent to magnetic transition metal oxides is a significant challenge~\cite{Freedman10}, particularly for gapless spin liquids. Beyond single-molecule magnets with purely local spin dynamics~\cite{Baker12}, spin degrees of freedom that are delocalized on stable organic molecules~\cite{Blundell04,Tamura09,Pratt11} or inorganic clusters~\cite{Sheckelton12}, can host collective electronic and magnetic phenomena controlled by interactions between magnetic molecules~\cite{Blundell04,Tamura09,Pratt11,Masuda08,Canevet10,Yamaguchi13}. Contrasting in the range and nature of interaction and disorder, different collective properties might be possible in such materials.

In this letter, we investigate the spin dynamics of \LZMO, an insulating compound where spin 1/2 carrying \MO\ clusters form a triangular lattice with dominant antiferromagnetic interactions~\cite{Sheckelton12}. Our inelastic neutron scattering data from powder specimens of \LZMO\ provide evidence for collective molecular magnetism. We observe a low-temperature spectrum from 0.2~meV to at least 2.5~meV compatible with the presence of valence-bonds entangling nearest- and next-nearest-neighbor \MO\ spin clusters and spectral weight corresponding to at most 1/3 of the spins.

The structure of \LZMO\ (resp. \ZMO) comprises \MO\ clusters organized in the hexagonal $ab$ plane of the $R\bar{3}m$ (resp.~$P6_3mc$) space-group. These planes are stacked along $c$ and separated by non-magnetic Zn and Li ions (resp.~Zn) with six (resp.~two) layers per unit-cell~\cite{Torardi85,Sheckelton12}. The \MO\ units maintain $C3v$ point-symmetry down to at least 12 K with an internal Mo--Mo distance $d_0\!=\!2.6$~\AA, and a greater distance $d_0^\prime\!=\!3.2$~\AA\ between Mo atoms of adjacent clusters~\cite{Sheckelton12}. This results in an effective triangular lattice of \MO\ units, Fig.~\ref{fig1}(a), with nearest-neighbor distance $d_1\!=\!5.8$~\AA\ between molecular centers. There are seven valence electrons (7$e$) per cluster in \LZMO\ and six (6$e$) in \ZMO. 

Electronic structure calculations for \MO(7$e$)~\cite{Sheckelton12} find a single unpaired electron delocalized in a non-degenerate molecular orbital. With the point-group symmetry of \MO, the wave-function is primarily concentrated on molybdenum with substantial weight on three oxygen atoms (marked $\ast$ in Fig.~\ref{fig1}(b)). This leads to a donut-shaped electron density of mean radius $\bar{r}\!\approx\!1.5$\AA\ [Fig.~\ref{fig1}(a)]. The large gap to higher-energy molecular levels suggests \MO(7$e$) carries spin $S=1/2$ [Fig.~\ref{fig1}(b)]. These theoretical predictions are corroborated by electron spin resonance (ESR)~\cite{Sheckelton13} and high-temperature susceptibility~\cite{Torardi85,Sheckelton12} measurements on powder-samples of \LZMO\ that are respectively described by an isotropic gyromagnetic tensor $g\!=\!1.9(1)$ (with upper bounds on anisotropy $g_\perp/g_\parallel\approx$ 0.8 to 1.2) and a matching effective moment $\mu_{\rm eff}\!=\!1.76(3) \mu_{\rm B}$~\cite{Suppl}, obtained by subtracting $\chi_0\!=\!-137(48)\times10^{-6}$emu.mol$^{-1}$.Oe$^{-1}$ from the data of Ref.~\cite{Sheckelton12}.  In contrast, the very small and temperature-independent susceptibility $|\chi_m|\!\approx\!|\chi_0|$  of \ZMO~\cite{Abe09,Sheckelton12} suggests that \MO(6$e$) is non-magnetic.

%% --------------------------------------------------------------------
\begin{figure}[t!]
\includegraphics[width=0.90\columnwidth]{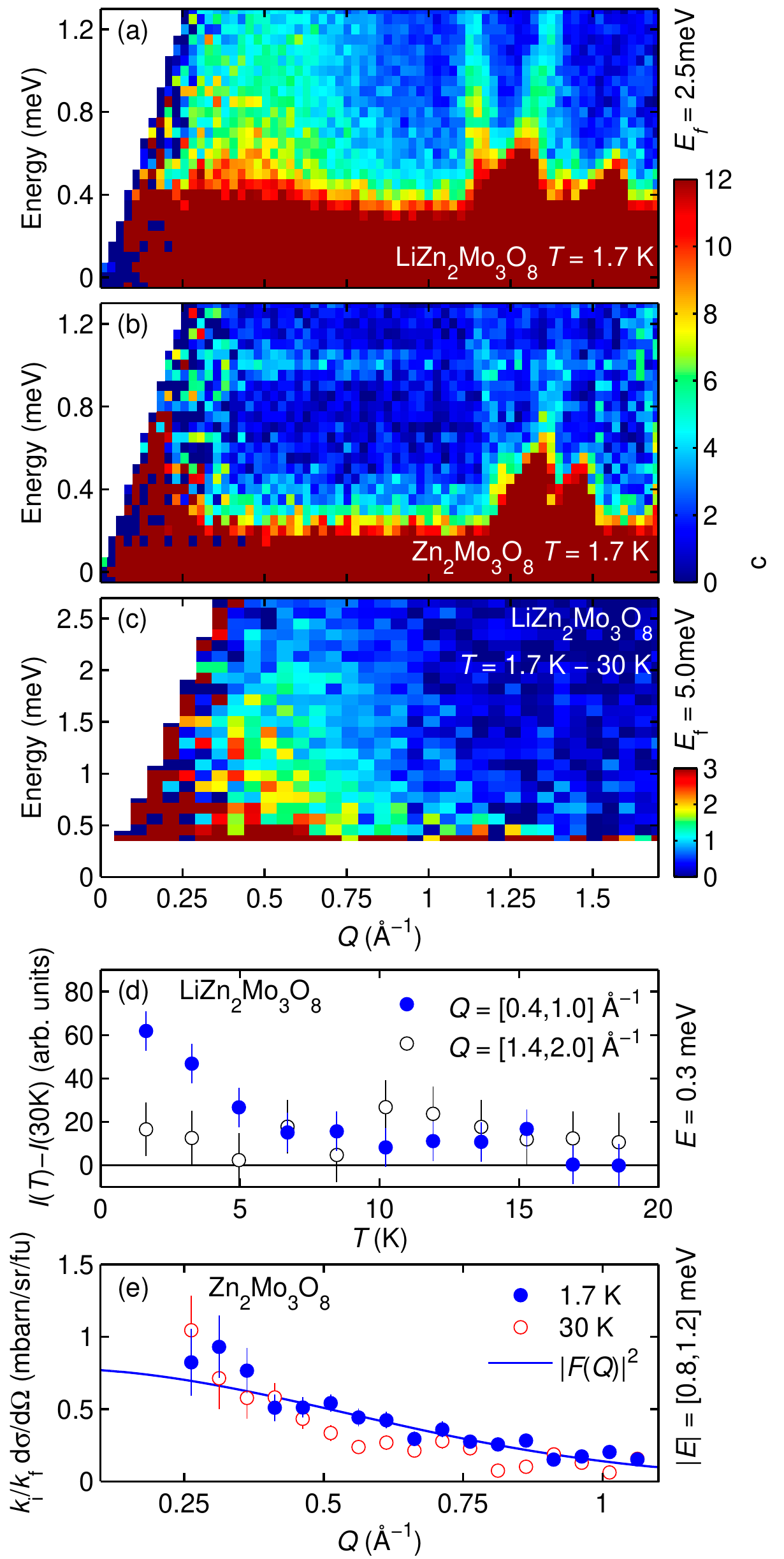}
\caption{(Color online) Neutron scattering cross-section for (a) \LZMO\ and (b) \ZMO\ at 1.7 K. The intensity is normalized to absolute units corresponding to mbarn meV$^{-1}$ sr$^{-1}$ per formula unit. (c) Temperature evolution from 30~K to 1.7~K in \LZMO\ ($E_f\!=\!5.0$~meV). (d) Temperature evolution at $E$=0.3 meV in \LZMO\ ($E_f\!=\!3.7$~meV) comparing low-$Q$ and high-$Q$ (full and open symbols, respectively). (e) Integrated intensity of the $E$=1.0 meV mode in \ZMO\ for 1.7 K and 30 K. The blue line is a comparison to the calculated $|F(Q)|^2$ for \MO(7$e$). Error bars represent one standard deviation. \vspace{-0.5cm}}
\label{fig2}
\end{figure}
%% --------------------------------------------------------------------

Antiferromagnetic interactions between \MO(7e) spins in \LZMO\ are indicated by a large negative Weiss constant $\Theta_W\!=\!-339(12)$~K~\cite{Suppl} for temperatures $T\!\geq\!100$~K. These may be primarily ascribed to super-exchange between adjacent clusters via short Mo--O--Mo (angles $97^\circ$ and $102^\circ$) paths. Below $T^*\!=\!96$~K, the susceptibility enters another effective Curie-Weiss regime with $\Theta_W^*\!=\!-27(6)$~K and $\mu_{\rm eff}^*\!=\!0.94(6)~\mu_{\rm B}$. This corresponds to an apparent loss of 71(4)\% ($\approx 2/3$) of the spins and was interpreted in Ref.~\cite{Sheckelton12} as a result of valence bond condensation. Despite a large $|\Theta_W|$, specific-heat shows no evidence for a transition to magnetic long-range order down to $0.1$~K but an upturn in the magnetic part of $C_p/T$ from $T\!\approx\!10$~K to 0.1~K~\cite{Sheckelton12}, consistent with a gapless magnetic excitation spectrum. 

To probe the corresponding magnetic excitations we carried-out inelastic neutron scattering experiments using the MACS spectrometer~\cite{Rodriguez08} at the NIST Center for Neutron Research and the ARCS spectrometer~\cite{Abernathy12} at the ORNL Spallation Neutron Source.  Powder samples of $^7$\LZMO\ ($m\!=\!18$ g) and \ZMO\ ($m\!=\!12$ g) were held in Al cans and cooled to $1.7$~K and $5.0$~K, respectively. The low energy regime ($\approx\Theta_W^*$) was explored on MACS using fixed final neutron energies $E_f\!=\!2.5$~meV, 3.7~meV and 5.0~meV and appropriate combinations of cooled Be and BeO filters before and after the sample to suppress higher-order contamination. Higher energies ($\approx\Theta_W$) were studied on ARCS configured with a fixed incident energy $E_i=154$~meV and a chopper frequency of 600~Hz. The corresponding Full Width at Half Maximum (FWHM) energy resolutions were 0.10~meV, 0.18~meV, 0.21~meV and 6.6~meV. Contributions from the empty cryostat were subtracted and the measured intensity was normalized to Bragg scattering from the sample~\cite{Xu13}.

We start with the low energy experiment for which the cross-section $\tilde{I}(\textit{Q},\textit{E})\equiv k_i/k_f \left({\rm d^2\sigma}/{\rm d}\Omega{\rm d}E_f\right)$ is plotted in Fig.~\ref{fig2}(a,b) as function of neutron energy-transfer $E \equiv \hbar\omega$ and momentum-transfer $\hbar Q \equiv \hbar |{\bf Q}|$. Besides elastic nuclear scattering, there is for \LZMO\ a broad plume of scattering extending from the elastic line up to the highest measured $E\!=\!1.3$~meV, concentrated at small $Q\!<\!1.0$~\AA$^{-1}$, and with a temperature-dependent characteristic wave-vector [Fig.~\ref{fig2}(a)]. There is no such signal for \ZMO\ in the same $Q$-range but instead, a weak flat mode at $E=1.01(1)$~meV the intensity of which decreases with $Q$ and vanishes by $Q\!\approx\!1.0$~\AA$^{-1}$ [Fig.~\ref{fig2}(b)]. For larger $Q\!>\!1.1$~\AA$^{-1}$, both samples display V-like ridges of intensity emerging from nuclear Bragg reflections. This spurious signal is temperature independent and results from incoherent elastic scattering from the monochromator or analyzer and a nuclear Bragg reflection from the sample. 

Our observations can be compared to the cross-section for inelastic magnetic neutron scattering associated with \MO\ spins, $\tilde{I}_m(Q,E)\!=\!r_0^2\left|\frac{g}{2}F(Q)\right|^2 2\,\tilde{S}(Q,E)$. Here $\tilde{S}(Q,E)$ is the dynamical spin correlation function, $F(Q)$ the spherically averaged form-factor for unpaired electrons in the sample and $r_0\!=\!0.539\!\times\!10^{-12}$~cm. Within the dipole approximation~\cite{Lovesey84} and assuming a quenched orbital contribution for \MO(7$e$), we obtain the spin-only from-factor $F({\bf Q})\!=\!\int{\rm d}^3{\bf r} \rho({\bf r}) e^{i{\bf Q}\cdot{\bf r}}$ from the unpaired electron density $\rho({\bf r})$ of Fig.~\ref{fig1}(a). 

%% --------------------------------------------------------------------
\begin{figure}[t]
\includegraphics[width=0.90\columnwidth]{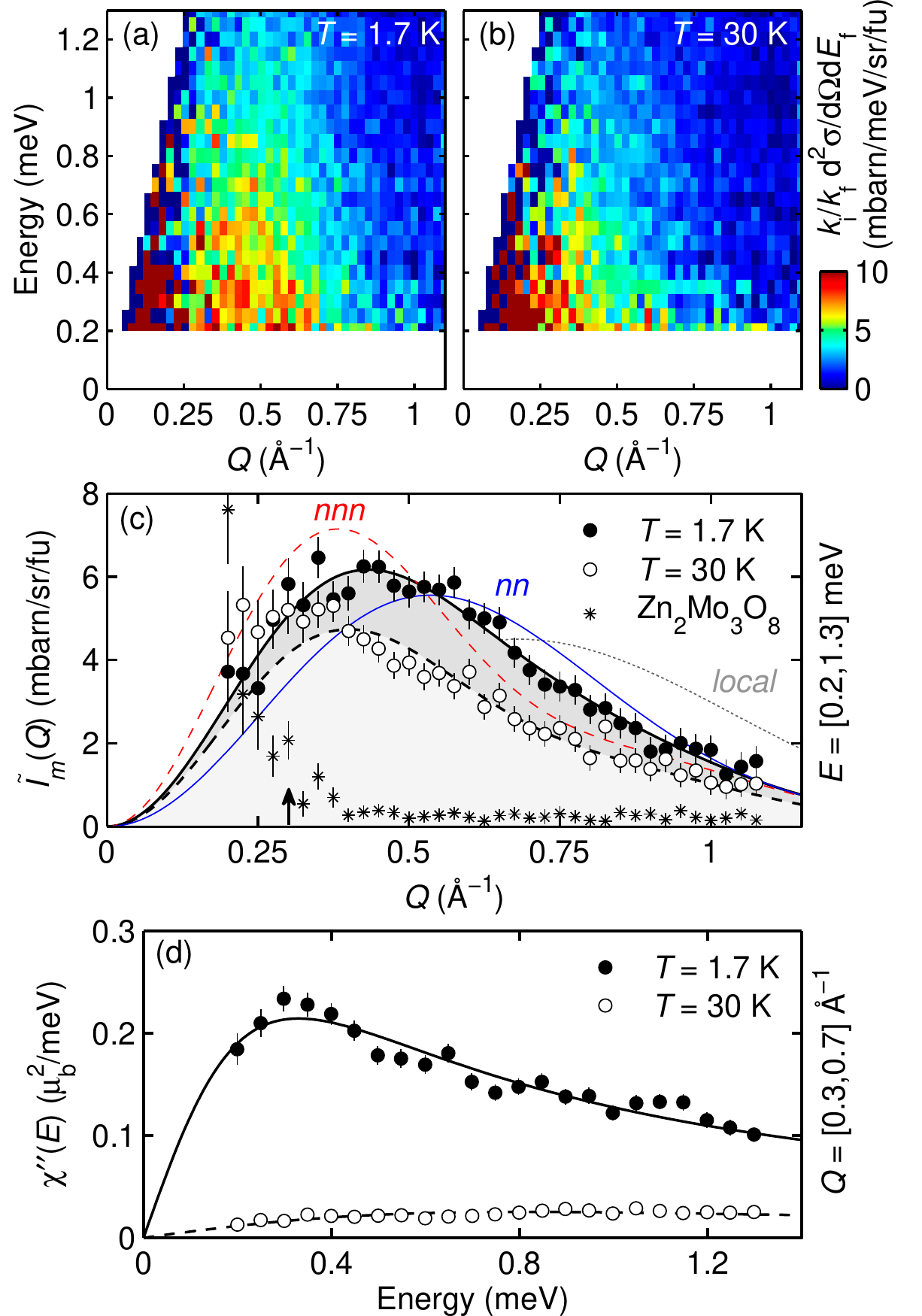}
\caption{(Color online)  Neutron scattering intensity in \LZMO\ ($E_f\!=\!2.5$~meV) at (a) 1.7 K and (b) 30~K corrected for the sample incoherent scattering. (c) Momentum dependence of $\tilde{I}_m(Q)$ for \LZMO\ compared to \ZMO\ ($0.2\!\leq\!{\it E}\!\leq\!0.8$~meV) revealing background contributions below $Q\!\approx\!0.3$\AA$^{-1}$ (black arrow). Fits to $\tilde{I}_{\rm vb}(Q)$ are indicated by thick black lines (solid 1.7~K, dashed 30~K). The thin lines are fits to $\tilde{I}_d(Q)$ at 1.7~K with $d_0\!=\!2.6$~\AA\ (dotted gray), $d_1\!=\!5.8$~\AA\ (solid blue) and $d_2\!=\!10.0$~\AA\ (dashed red). (d) Energy dependence of $\chi^{\prime\prime}(E)$ at 1.7 K (full symbols) and 30 K (open symbols). Solid and dashed lines are fits to a relaxation response. Error bars represent one standard deviation. \vspace{-0.5cm} }
\label{fig3}
\end{figure}
%% --------------------------------------------------------------------

The spherically averaged squared amplitude $|F({Q})|^2$ decreases with increasing $Q$ and drops to 10\% of its initial value by $Q\!\approx\!1.0$~\AA$^{-1}$ [Fig.~\ref{fig1}(b)]. This resembles the trend observed experimentally for small $Q$ in Fig.~\ref{fig2}(a,b), suggesting that both the broad signal in \LZMO\ and the flat excitation in \ZMO\ have a magnetic origin. For \LZMO\ this is reinforced by the temperature evolution of the signal, determined by subtracting $30$~K data from lower temperature measurements. Upon cooling to $1.7$~K, the intensity increases for $Q\!<\!1.0$~\AA$^{-1}$ and from the elastic line up to at least $E\!=\!2.5$~meV [Fig.~\ref{fig2}(c)]. A more detailed temperature dependence focusing on $E=0.3$~meV reveals a substantial decrease of the $Q\!<\!1.0$~\AA$^{-1}$ signal from $T=1.7$~K to $T\!\approx\!10$~K while the $Q\!>\!1.4$~\AA$^{-1}$ is $T$-independent [Fig.~\ref{fig2}(d)]. 

In contrast, the signal observed in \ZMO\ [Fig.~\ref{fig2}(b)] consists of a weak resonant mode with integrated intensity  that follows $|F(Q)|^2$ remarkably well, particularly for $0.4\!<\!Q\!<\!1.0$~\AA$^{-1}$ [Fig.~\ref{fig2}(e)].  This flat mode carries a temperature independent spectral weight corresponding to $\approx\!10$\% of that observed in \LZMO\ [Fig.~\ref{fig2}(a)] or about 3\% of that expected from one $S\!=\!1/2$ per \MO\ cluster. We associate this scattering with a local intra-molecular excitation of \MO($6e$) that validates the general trend of our \textit{ab-initio} predictions for the form-factor [Fig.~\ref{fig1}(b)]. 

For a more quantitative understanding, we isolated the inelastic magnetic scattering contribution, $\tilde{I}_m(Q,E)$, by subtracting the sample elastic nuclear scattering. Specifically, the elastic incoherent lineshape observed in \ZMO\ for ${\it E}\!<\!0.8$~meV and $0.3\!\le\!Q\!\le\!1.1$~\AA$^{-1}$ was scaled to the $\tilde{I}(Q,E\!=\!0)$ intensity of \LZMO\ and subtracted [Fig.~\ref{fig3}(a,b)]. 

The momentum dependence of the resulting intensity, $\tilde{I}_m(Q)=\int {\rm d}E\, \tilde{I}_m(Q,E)$, was extracted by integrating over $0.2\!<\!{\it E}\!<\!1.3$~meV. A peak is observed for $Q\!=\!0.41(2)$\AA$^{-1}$ at $1.7$~K which shifts to a lower $Q\!=\!0.35(2)$\AA$^{-1}$ upon warming to $30$~K [Fig.~\ref{fig3}(c)]. This indicates the signal is a collective excitation of the \MO(7e) spins rather than a local intra-molecular excitation. The latter would peak at a much higher $Q$, see the dotted line in Fig.~\ref{fig3}(c), and be temperature independent. 

We modeled $\tilde{I}_m(Q)$ using the powder-averaged equal-time structure factor of a valence bond, $\tilde{I}_d(Q) \!\propto\!\left|F(Q)\right|^2\left[1\!-\!{\sin(Qd)}/{(Qd)}\right]$ with $d$ the distance between antiferromagnetically interacting spins.  Fits to the data in the range $0.3\!\leq\!Q\!\leq\!1.1$ \AA$^{-1}$  with variable $d$ (not shown) yield an effective $d^*\!=\!7.4(2)$~\AA\ at $1.7$~K indicating temperature-dependent correlations that are longer-ranged than the nearest-neighbor spacing of $5.8$~\AA. Fixing $d$ to $d_1$ (blue solid line) or to the next-nearest-neighbor distance $d_{2}\!=\!10.0$~\AA\ (red dashed line) does not yield a satisfactory fit to the $1.7$~K data [Fig.~\ref{fig3}(c)]. A much better fit is obtained by allowing the superposition of valence bonds for several near neighbors $\tilde{I}_{\rm vb}(Q) = r_0^2/6\left|F(Q)\right|^2 S_{\rm vb}(Q)$ where $S_{\rm vb}(Q)=\sum_{i=1}^2 m^2_i\left[1-{\sin(Qd_i)}/{(Qd_i)}\right]/\mu^2_{\rm B}$ with $m^2_i$ the squared moment per formula unit entangled in a valence-bond $d_i$. Fits to this model, shown in Fig.~\ref{fig3}(c) for $1.7$~K (solid bold line) and $30$~K (dashed bold line), yield $m^2_1\!=\!0.10(1)\mu_{\rm B}^2$ ($55(9)\%$ of $\sum_im^2_i$) and $m^2_2\!=\!0.08(1)\mu_{\rm B}^2$ ($45(7)\%$) at 1.7~K, and $m^2_1\!=\!0.03(1)\mu_{\rm B}^2$ ($27(11)\%$) and $m^2_2\!=\!0.09(1)\mu_{\rm B}^2$ ($73(15)\%$) at 30~K. These results are independent of the background subtraction within error bars~\cite{Suppl}. Including a third neighbor distance $d_3$ does not significantly improve the fits and strongly depends on the background subtraction. While limited to low energy, our findings are consistent with the superposition of valence-bonds involving first and second nearest-neighbors at 1.7~K. The structure factor shifts to lower $Q$ upon warming to 30~K, a phenomenology also observed in the Kagome quantum spin liquid Kapellasite~\cite{Fak12}.

The energy dependence of the signal was analyzed through the imaginary part of the dynamical susceptibility ${\chi}^{\prime\prime}(Q,E)\!=\!\pi (g\mu_{\rm B})^2 (1 - e^{-\beta{\it E}})\tilde{S}(Q,E)$. The momentum-integrated susceptibility ${\chi}^{\prime\prime}(E)$ is shown in Fig.~\ref{fig3}(d) for $0.3\!<\!{\it E}\!<\!0.7$~meV. There is no discernible gap or resonance and the data are well described by a relaxation response ${\chi}^{\prime\prime}(E)\!=\! {\chi}^\prime {\it E} \Gamma / ({\it E}^2+\Gamma^2)$ with relaxation-rate $\Gamma\!=\!0.36(1)$~meV at 1.7~K and $\Gamma\!=\!0.91(5)$~meV at 30~K. The temperature dependence of $\Gamma$ and its similarity in magnitude to $|\Theta_W^*|\!=\!2.3(5)$~meV again point to a collective phenomenon.

%% --------------------------------------------------------------------
\begin{figure}[t!]
\includegraphics[width=0.80\columnwidth]{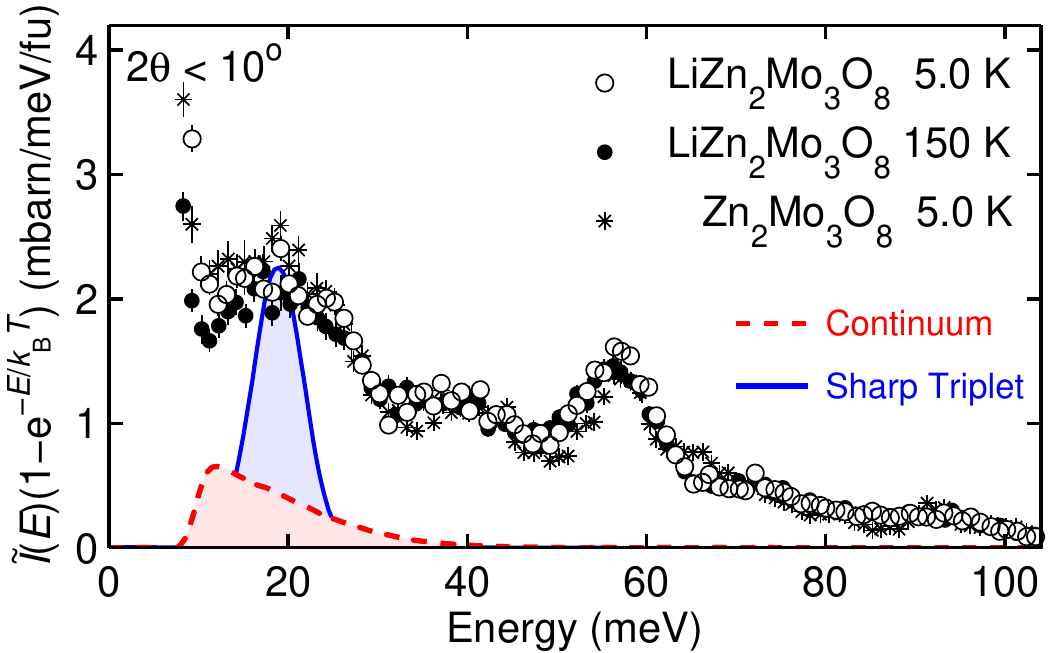}
\caption{(Color online) Energy dependence of the neutron scattering intensity ($E_i$=154~meV) at low scattering angles ($2\theta\!<\!10^{\circ}$) for \LZMO\ at 5~K (open circles) and 150~K (black circles) and \ZMO\ at 5~K (black stars). The solid blue and dashed red lines are predictions for a sharp triplet excitation ($\Delta=J_{\rm av}$) and continuum  ($\Delta=J_{\rm av}/2$, bandwidth $2J_{\rm av}$), respectively. Error bars represent one standard deviation. \vspace{-0.5cm}}
\label{fig4}
\end{figure}
%% --------------------------------------------------------------------

The inelastic spectral weight per formula-unit $m^2\!=3 \mu_{\rm B}^2 \!\int\!\!\!\int\! Q^2 [g^2\tilde{S}(Q,E)] {\rm d}Q{\rm d}E/\!\int\! Q^2{\rm d}Q $ can be directly compared to $\mu^{*2}_{\rm eff}\!=\!0.94(6)~\mu_{\rm B}^2$ derived from bulk susceptibility data for $20\!<\!T\!<\!90$~K and associated with $\approx\!1/3$ of the spins. Accounting for the intensity on the neutron energy gain side and integrating over the range $0.2\!<\!E\!<\!1.3$~meV and $0.3\!<\!Q\!<\!1.1$~\AA$^{-1}$ yields $m^2\!=\!0.23(4)\,\mu^2_B$ at 1.7~K and $m^2\!=\!0.25(7)\,\mu^2_B$ at 30~K. This agrees well with $\sum_{i}\!m^2_i$ obtained above and yields $m^2\!\approx\!0.27\mu^{*2}_{\rm eff}$ (and thus only $\approx\!0.08 \mu^{2}_{\rm eff}$). The shortfall is attributed to the limited range of $E$-integration. To overcome this, we modeled the dynamic structure factor as $g^2\tilde{S}_{\rm th}(Q,E)\!=\!S_{\rm vb}(Q)f(E)$ where $f(E)\!=\! \mathcal{C}(T) \theta(E^*-|E|) \Gamma E/[(\Gamma^2+E^2)(1-e^{-\beta E})]$ with $E^*$ a high-energy cutoff and $\mathcal{C}(T)$ a dimensionless prefactor to ensure normalization $\int\!\!f(E){\rm d}E\!=\!1$. With the parameters from the best fits, the reconstructed spectral weight extrapolates to $\mu^{*2}_{\rm eff}$ for a cutoff energy $E^*\!\approx\!11$~meV, indicating that a maximum of 1/3 of the spins participate in this low-energy signal.

Searching for higher energy spectral weight, Fig.~\ref{fig4} presents the energy dependence of the neutron scattering intensity $\tilde{I}(E<100~$meV) at low scattering angles ($2\theta\!<\!10^{\circ}$). The scattering is dominated by broad features centered around 20~meV, 42~meV and 57~meV that are common to \LZMO\ and \ZMO. With an intensity that increases with $Q$~\cite{Suppl} and varies with $T$ in accordance with the Bose factor, they are associated with acoustic and optical phonons. Their large density-of-states at low-energies hints them to \MO\ clusters, thus confirming the stability of these molecular units. In contrast, no magnetic signal is apparent.

Using an average exchange interaction $J_{\rm av}\!=\!3k_{\rm B}\Theta_{W}/6S(S+1)\!\approx\!19.5$~meV, we modeled the energy dependence of $\tilde{I}_{\rm m}(Q,E)$ for a singlet to triplet transition with distance $d_1$ and integrated spectral-weight $m^2=\mu_{\rm eff}^{2}-\mu_{\rm eff}^{*2}=2.20(16)\mu_{\rm B}^2$ in two cases: a resolution-limited transition at $\Delta\!=\!J_{\rm av}$ and a gapped continuum with $\Delta\!=\!J_{\rm av}/2$ and bandwidth $2J_{\rm av}$ [Fig.~\ref{fig4}]. While the sharp excitation, if present, would be clearly visible at low scattering angles, the continuum mostly develops out of the kinematic range of our experiment~\cite{Suppl} with a very weak accessible intensity, representing a fraction of the phonon contribution. While we cannot exclude a sharp excitation with $\Delta>30$~meV, the analysis suggests the missing spectral weight may as well hide in a weak $10$--$40$~meV continuum obscured by phonon scattering.

The magnetism of \LZMO\ differs radically from that of the spin-1/2 triangular lattice Heisenberg antiferromagnet, which has long-range spin  order~\cite{Huse88,Jolicoeur89}. Along with previously published thermo-magnetic data~\cite{Sheckelton12}, our neutron scattering data indicate a quantum fluctuating low temperature state with low-energy excitations carrying at most a third of the spectral weight expected from the high-temperature effective moment.  The low-energy spin correlations are compatible with valence-bonds involving first and second nearest neighbor \MO\ clusters. The apparently gapless spectrum indicates disorder or longer-range valence bonds than directly detectable here. Correspondingly, the absence of detectable sharp excitation at higher energies is consistent with a disordered valence bond solid or a resonating valence bond state. 

Potentially relevant deviations from the Heisenberg model that may explain these results include exchange anisotropies, longer-range interactions, multi-spin exchange, and magneto-elasticity. Recently, a mechanism based on the cooperative rotation of \MO\ clusters resulting in an emergent honeycomb lattice has been proposed for \LZMO~\cite{Flint13}. Another interesting direction is Schwinger boson mean-field theory~\cite{Messio11}. It may be possible to distinguish between these scenarios through lower $Q$ inelastic data or by examining the magnetic field dependence of the present data. 

We thank R. Flint, P. A. Lee, O. Tchernyshyov and Y. Wan for discussion and Y. Qiu for assistance with the MACS experiment and data treatment. The work at IQM was supported by the US Department of Energy, office of Basic Energy Sciences, Division of Material Sciences and Engineering under grant DE-FG02-08ER46544. This work utilized facilities supported in part by the National Science Foundation under Agreement No. DMR-0944772. Research at Oak Ridge National Laboratory's Spallation Neutron Source is sponsored by the US Department of Energy, office of Basic Energy Sciences, Scientific User Facilities Division.

% -------------------------------------------------------------------
% Supplementary materials
% -------------------------------------------------------------------
\clearpage
\onecolumngrid
\appendix
%\doublespace
\setcounter{figure}{0}
\setcounter{table}{0}
\renewcommand{\thefigure}{S\arabic{figure}}
\renewcommand{\thetable}{S\arabic{table}}
% -------------------------------------------------------------------

\begin{center}
{\large \bf Supplementary online material for  ``Molecular Quantum Magnetism in \LZMO''}
\end{center}

{\noindent \bf 1. Curie-Weiss analysis of the bulk susceptibility} \\

\begin{figure}[h!]
\includegraphics[width=0.55\columnwidth]{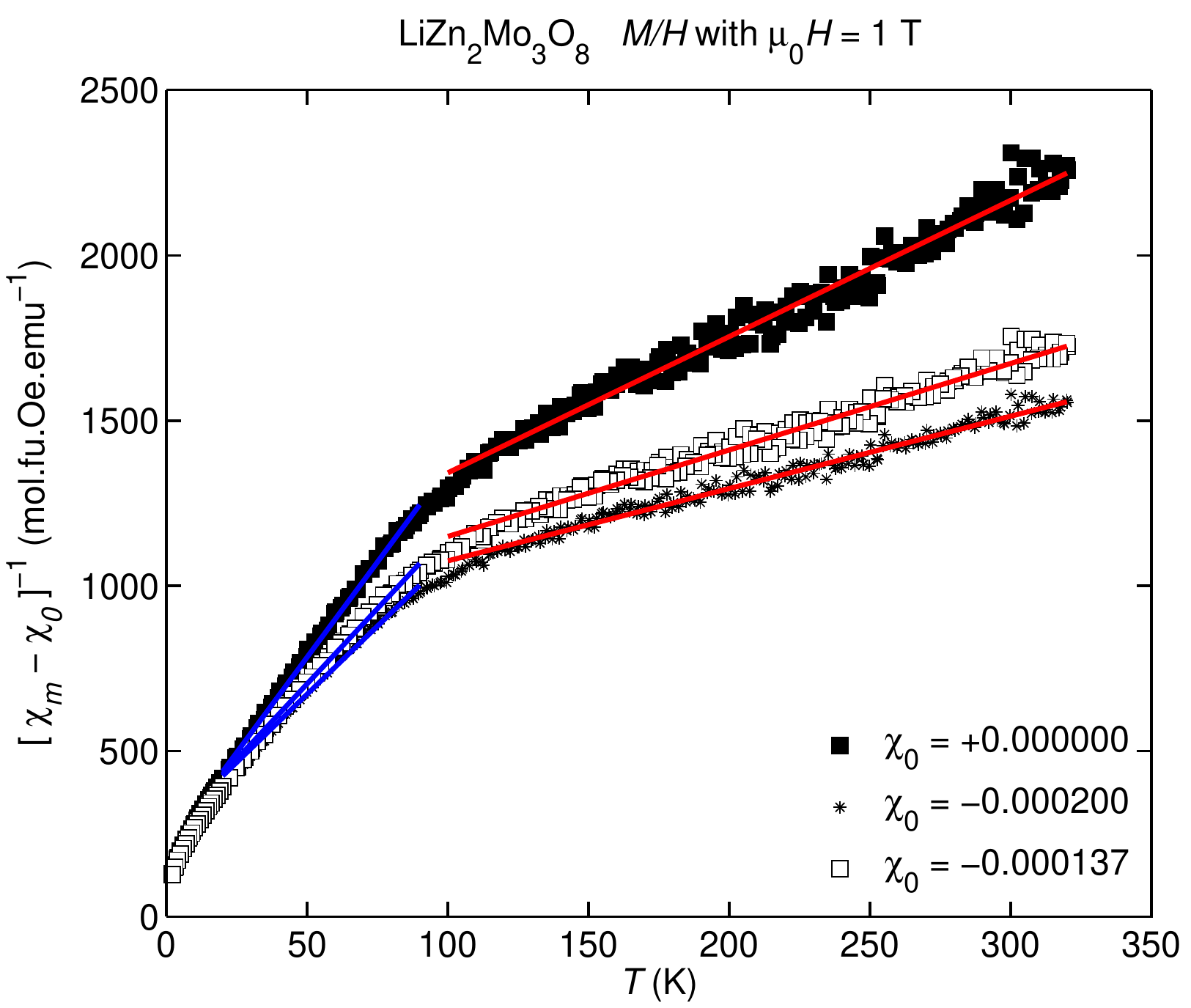}
\caption{Inverse molar susceptibility $[\chi_m(T)-\chi_0]^{-1}$ of \LZMO: taken from Ref.~16 (black squares), corrected for a small negative $\chi_0$ obtained by fitting (white squares) and corrected for an empirical estimate of $\chi_0$ (black stars). The red and blue lines are representative Curie-Weiss fits to the $T>100$~K and $T<90$~K regimes, respectively.}
\label{figs1}
\end{figure}

\begin{figure}[t!]
\includegraphics[width=0.75\columnwidth]{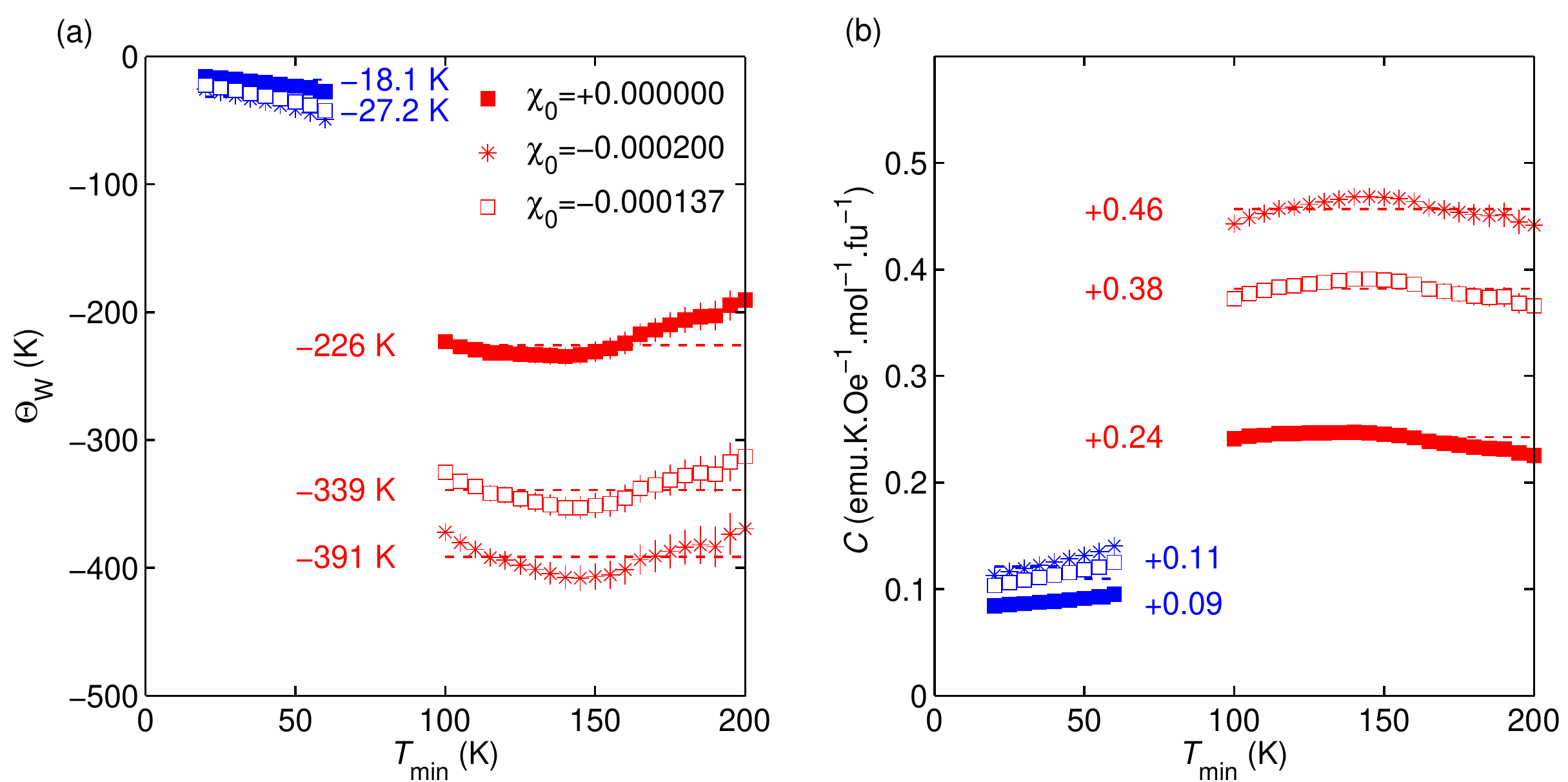}
\caption{Curie-Weiss analysis as a function of the fitting range and value of the subtracted $\chi_0$: (a) Weiss temperature $\Theta_{W}$ and (b) Curie constant $C$. At low-temperature (blue marks) the fit is performed in the range between $T_{\rm min}$ and 90~K with $T_{\rm min}\in [20,60]$~K. At high-temperature (red marks) the fit is performed in the range between $T_{\rm min}$ and 330~K with $T_{\rm min}\in [100,200]$~K. The dashed lines indicate the average result of fits over the range of $T_{\rm min}$. }
\label{figs2}
\end{figure}

In  Ref.~\cite{Sheckelton12}, diamagnetic contributions ($\chi_0$) to the susceptibility of \LZMO\ were determined using an equimolar amount of the non-magnetic sample \ZMO\ mounted on the same sample holder. This approach has the advantage of being independent of the data analysis procedure, but has limitations related to the stability of the instrumentation. In this work we determine $\chi_0$ by fitting the data of Ref.~16 for temperatures between $T_{\rm min}$ and 330~K with $T_{\rm min}\!\in\![100,250]$~K. The fit yields $\chi_0=-137(48)\times10^{-6}$~(emu/mol\,fu\,Oe) in excellent agreement with the empirical estimate $\chi_0=-200\times10^{-6}$~(emu/mol\,fu\,Oe). The errorbar represents the standard deviation of $\chi_0$ over the range of $T_{\rm min}$. The inverse susceptibility of \LZMO\ corrected for $\chi_0$ is presented in Fig.~\ref{figs1} along with representative non-linear least-squares fits of the two distinct $T>100$~K and $T<90$~K Curie-Weiss regimes.  In Fig.~\ref{figs2}, the fit results are presented versus fitting range and $\chi_0$. Statistically averaged values over the range of $T_{\rm min}$ are summarized in Tab.~\ref{tab1} along with comparison to predictions for $S=1/2$ spins with $g=2.0$ (free electron), $g=1.9(1)$ (obtained by ESR in Ref.~22) or $g=1.6$ (obtained in Ref.~16) and $N=1$ or $N=1/3$ spins per formula-unit.  The results are $\Theta_{W}=-339(12)~{\rm K}$, $\mu_{\rm eff}^2 =3.08(6)\mu_{\rm B}^2$  for $T>100~{\rm K}$ and $\Theta^*_{W}=-27(6)~{\rm K}$, $\mu_{\rm eff}^{*2}=0.88(6)\mu_{\rm B}^2$  for  $T<90~{\rm K} $.

%% --------------------------------------------------------------------
%% --------------------------------------------------------------------

%% --------------------------------------------------------------------
\begin{table}[b!]
	\centering
		\begin{tabular}{|l|lllll|lllll|}
		\hline\hline
    $\chi_0$     &(a) & $\Theta_{W}$   & $C$	 & $\mu^2_{\rm eff}$   & $\mu_{\rm eff}$
  	             &(b) & $\Theta_{W}^*$ & $C^*$ & $\mu^{*2}_{\rm eff}$ & $\mu^*_{\rm eff}$ \\ \hline
   -0.00000(0)   & $T\geq100$ K \ \ & -226(15) & 0.24(1) & 1.96(6) & 1.40(4)  
  				       & $T\leq90$  K \ \ & -18(7)   & 0.09(1) & 0.70(3) & 0.84(3)      \\		
   -0.00137(45)  & $T\geq100$ K \ \ & -339(12) & 0.38(1) & 3.08(6) & 1.76(3)  
  				       & $T\leq90$  K \ \ & -27(6)   & 0.11(1) & 0.88(6) & 0.94(6)      \\ 
   -0.00200(0)   & $T\geq100$ K \ \ & -391(12) & 0.46(1) & 3.68(6) & 1.91(4)  
    			       & $T\leq90$  K \ \ & -32(7)   & 0.12(1) & 0.97(7) & 0.99(7)	  	\\ \hline
    $g=1.60$ &$N=1$ &	-- & 0.24 & 1.92 & 1.39 & $N=1/3$ & -- & 0.08 & 0.64 & 0.80 \\ 		
    $g=1.90$ &$N=1$ & -- & 0.33 & 2.70 & 1.64 & $N=1/3$ & -- & 0.11 & 0.90 & 0.95 \\
    $g=2.00$ &$N=1$ & -- & 0.37 & 3.00 & 1.73 & $N=1/3$ & -- & 0.12 & 1.00 & 1.00 \\	\hline\hline 		
 		\end{tabular}
		\caption{Curie-Weiss analysis of the susceptibility of \LZMO\ and comparison with predictions for
		 paramagnetic spin 1/2: (a) above $T$=100~K and (b) below $T$=90~K with $\chi_0$ in units of (emu/mol\,fu\,Oe), $C$ in units of (emu\,K/Oe\,mol\,fu), $\Theta_{W}$ in K, and $\mu_{\rm eff}$ in $\mu_{\rm B
		}$.}
\label{tab1}
\end{table}
%% --------------------------------------------------------------------

%% --------------------------------------------------------------------
%% --------------------------------------------------------------------

%% --------------------------------------------------------------------
\begin{table}[h!]
	\centering
		\begin{tabular}{|cc|l|c|ccccc|c|}
		\hline\hline
		     &  & Valence Bonds         & \hspace{0.4cm}  $\chi^2$ \hspace{0.4cm}  
				  & $m^2_1/\mu_{\rm B}^2$ \ \ \ \ & $m^2_2/\mu_{\rm B}^2$  \ \ \ \ 
				  & $m^2_3/\mu_{\rm B}^2$ \ \ \ \ & $m^2_0/\mu_{\rm B}^2$  \ \ \ \  
					& $m^2/\mu_{\rm B}^2$  \ \ \ \  &  \\ \hline\hline
	   (a) raw & 1.7 K & $d_1$ + $d_2$ + $d_3$ & 0.63   & 0.10(1) & 0.08(1) & 0.00(2) &    --   &    --  &  \\
		     &       & $d_1$ + $d_2$         & 0.61   & 0.10(1) & 0.08(1) & --      &    --   &    --  & 
				$\Longleftarrow$ \\
		     &       & $d_2$                 & 4.07   &  --     & 0.16(3) & --      &    --   &    --  &  \\
				 &       & $d_1$                 & 3.76   & 0.19(3) &  --     & --      &    --   &    --  &  \\
				 &       & $d_0$                 &21.74   &  --     &  --     & --      & 0.47(1) &    --  &  \\ 
				 &       & $d=7.4(2)$~\AA        & 1.11   &  --     &  --     & --      &    --   & 0.17(1)& \\ \hline			
	  (c) sub & 1.7 K & $d_1$ + $d_2$ + $d_3$ & 0.58   & 0.11(1) & 0.07(1) & 0.00(0) &    --   &    --  & \\
		     &       & $d_1$ + $d_2$         & 0.54   & 0.11(1) & 0.07(1) & --      &    --   &    --  & 
				$\Longleftarrow$ \\
		     &       & $d_2$                 & 3.57   &  --     & 0.15(3) & --      &    --   &    --  & \\
				 &       & $d_1$                 & 2.23   & 0.18(3) &  --     & --      &    --   &    --  & \\
				 &       & $d_0$                 &15.64   &  --     &  --     & --      & 0.44(1) &    --  & \\
				 &       & $d=11.0(2)$~\AA        & 1.41   &  --     &  --     & --      &    --   & 0.12(1)& \\ 
		\hline\hline										
    (b) raw & 30  K & $d_1$ + $d_2$ + $d_3$ & 0.64   & 0.04(1) & 0.00(0) & 0.09(1) &    --   &    --  & 
		$\Longleftarrow$ \\
		     &       & $d_1$ + $d_2$         & 0.96   & 0.03(1) & 0.09(1) & --      &    --   &    --  & \\
		     &       & $d_2$                 & 1.41   &  --     & 0.12(2) & --      &    --   &    --  & \\
				 &       & $d_1$                 & 5.84   & 0.13(0) &  --     & --      &    --   &    --  & \\
				 &       & $d_0$                 &18.31   &  --     &  --     & --      & 0.32(1) &    --  & \\
				 &       & $d=7.4(2)$~\AA        & 1.11   &  --     &  --     & --      &    --   & 0.17(1)& \\ \hline
    (d) sub & 30  K & $d_1$ + $d_2$ + $d_3$ & 0.50   & 0.04(1) & 0.05(3) & 0.03(3) &    --   &    --  & 
		    $\Longleftarrow$ \\
		     &       & $d_1$ + $d_2$         & 0.50   & 0.03(1) & 0.08(1) & --      &    --   &    --  & 
				$\Longleftarrow$  \\
		     &       & $d_2$                 & 0.92   &  --     & 0.11(2) & --      &    --   &    --  & \\
				 &       & $d_1$                 & 3.15   & 0.12(0) &  --     & --      &    --   &    --  & \\
				 &       & $d_0$                 &11.59   &  --     &  --     & --      & 0.28(1) &    --  & \\ 
				 &       & $d=8.8(5)$~\AA        & 1.11   &  --     &  --     & --      &    --   & 0.10(1)& \\ \hline										
 		\end{tabular}
		\caption{Valence-bond fit results associated with the panels of Fig.~\ref{figs3}. The double arrows indicate 
		the best fits based on the values of $\chi^2$.
		\label{tabs2}}
\end{table}

{\noindent \bf 2. Valence-bond fits} \\

\begin{figure}[h!]
\includegraphics[width=0.85\columnwidth]{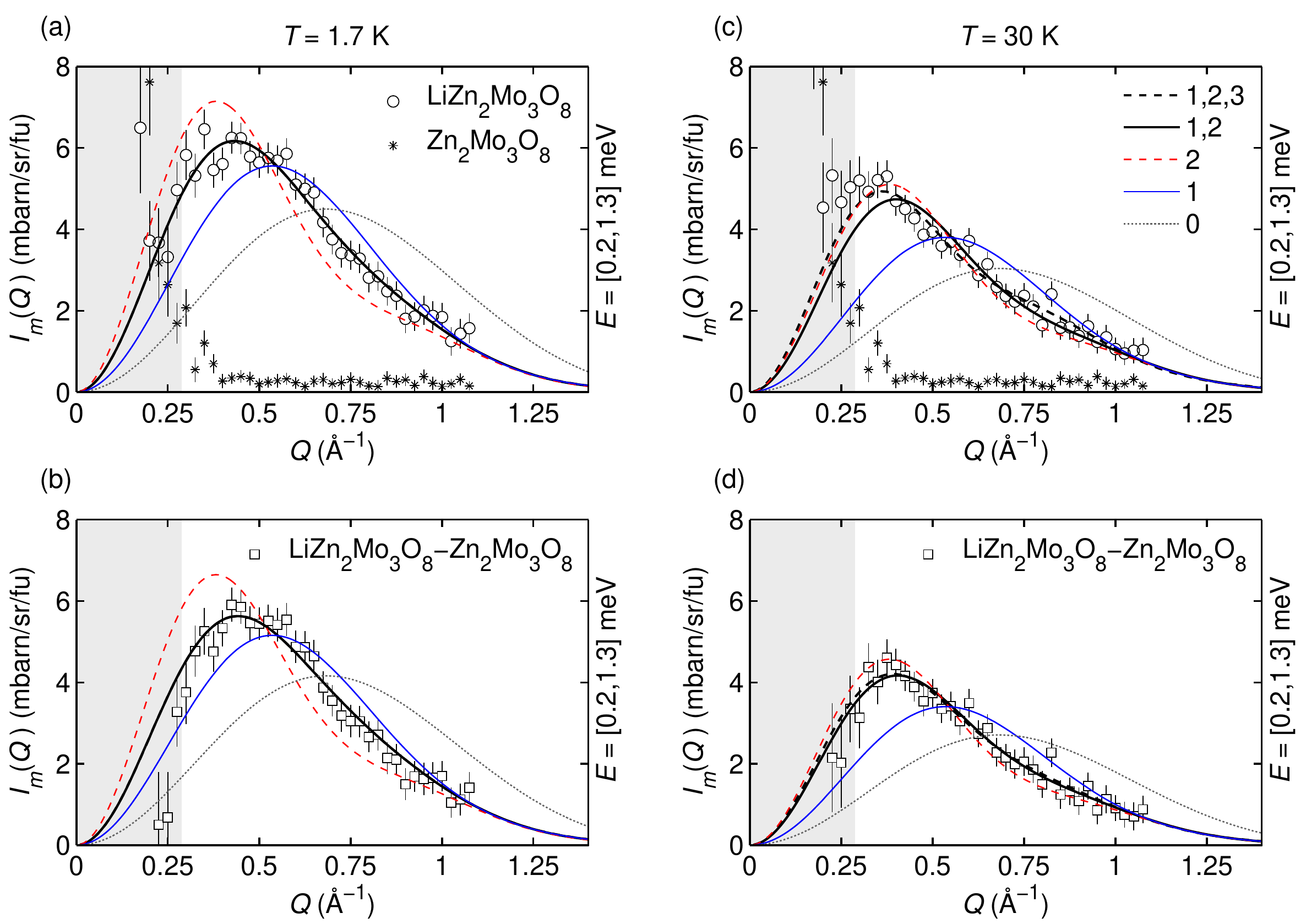}
\caption{Momentum dependence of the inelastic magnetic scattering intensity $\tilde{I}_{\rm m}(Q)$ for:
\LZMO\ (white circles) with $0.2\!<\!{\it E}\!<\!1.3$~meV and $T=1.7$~K or $T=30$~K, \ZMO\ (black stars) with $0.2\!<\!{\it E}\!<\!0.8$~meV and \LZMO\ subtracted for the \ZMO\ background (white squares). Lines are fits to different combinations of $d_i$ valence-bonds ($i=0,1,2,3$) for $0.3<Q<1.1$~\AA$^{-1}$.}
\label{figs3}
\end{figure}

Details of valence-bond fits to $\tilde{I}_m(Q)$ are shown in Fig.~\ref{figs3} with the cut-off $Q<0.3$~\AA$^{-1}$ indicated by the gray area.  The raw \LZMO\ data [Fig.~\ref{figs3}(a,c)] is plotted along with an empirical background obtained for \ZMO\ in the range $0.2\!<\!{\it E}\!<\!0.8$~meV. Background subtracted data  [Fig.~\ref{figs3}(b,d)] are not shown in the main text because the background is obtained on a different and limited energy range compared to the signal. Fit results are plotted using the same convention as in the main text and results summarized in Tab.~\ref{tabs2} using fixed distances $d_0$, $d_1$, $d_2$, $d_3$ and free $d$.

%% --------------------------------------------------------------------
%% --------------------------------------------------------------------

\clearpage
{\noindent \bf 3. High-energy neutron scattering spectra} 

%% --------------------------------------------------------------------
\begin{figure}[h!]
\includegraphics[width=0.99\columnwidth]{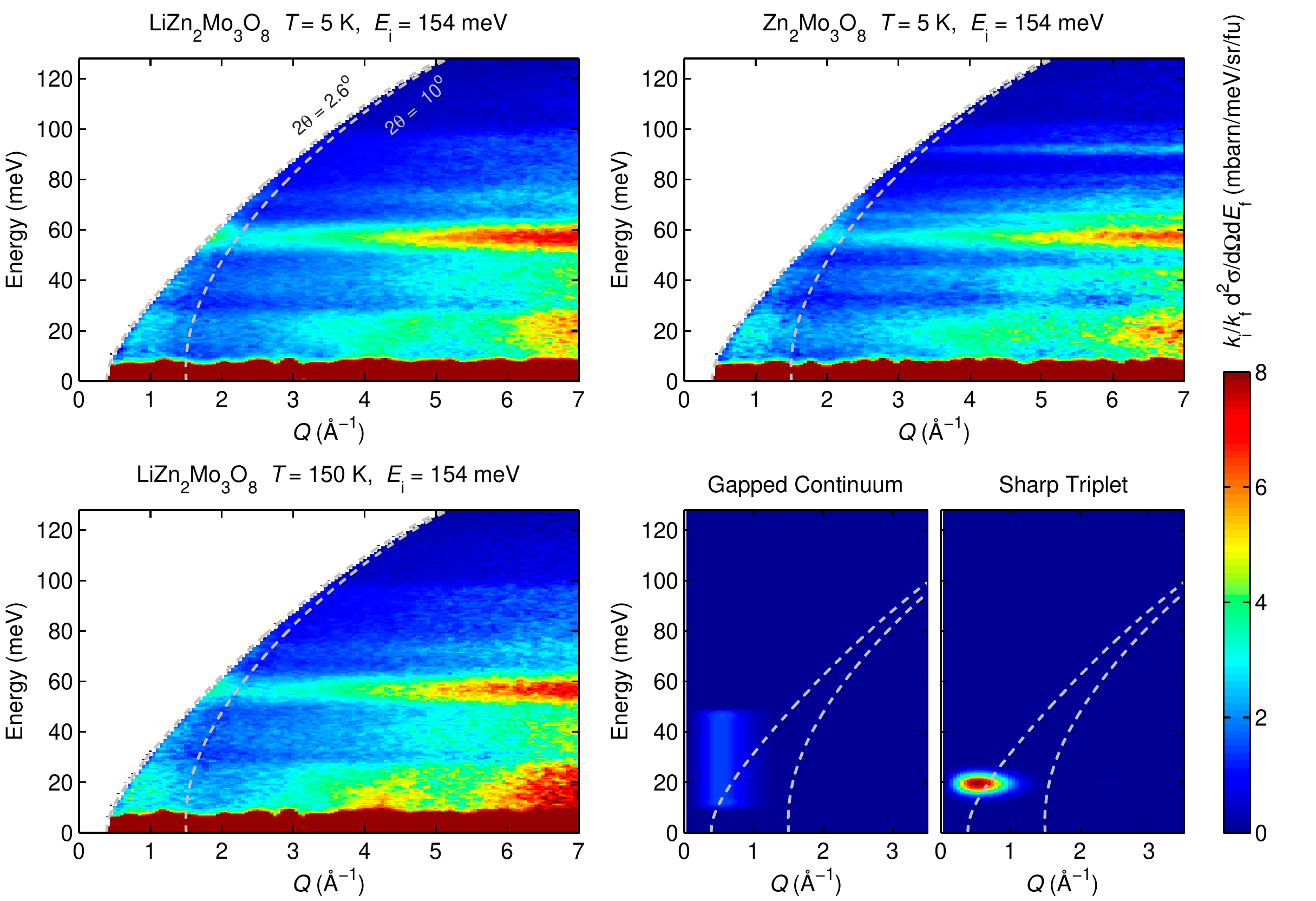}
\caption{Neutron scattering intensity obtained on ARCS ($E_i=154$~meV) for \LZMO\ at 5~K and 150~K and \ZMO\ at 5~K compared to predictions for gapped continuum and sharp triplet excitations. The dashed lines indicate the region of low scattering angles shown in the energy dependence of Fig.~\ref{fig4}.}
\label{figs4}
\end{figure}
%% --------------------------------------------------------------------


\begin{thebibliography}{99}

\bibitem{Anderson73} 
P.~W.~Anderson, Mater. Res. Bull. {\bf 8}, 153-160 (1973).

\bibitem{Lee08} 
%From high temperature superconductivity to quantum spin liquid: progress in strong correlation physics
P. A. Lee, Reports on Progress in Physics {\bf 71}, 012501 (2008).

\bibitem{Stone03} 
%Extended Quantum Critical Phase in a Magnetized Spin-1/2 Antiferromagnetic Chain. 
M. B. Stone, D. H. Reich, C. Broholm, K. Lefmann, C. Rischel, C. P. Landee, and M. M. Turnbull,
\prl, {\bf 91}, 037205 (2003).

\bibitem{Lake10} 
% Confinement of fractional quantum number particles in a con-densed-matter system‘
B. Lake, A. M. Tsvelik, S. Notbohm, D. A. Tennant, T. G. Perring, M. Reehuis, C. Sekar, G. Krabbes, and B. B\"uchner,
\natphys {\bf 6}, 50-55 (2010).

\bibitem{Coldea10} 
%Quantum Criticality in an Ising Chain: Experimental Evidence for Emergent E8 Symmetry
R. Coldea, D. A. Tennant, E. M. Wheeler, E. Wawrzynska, D. Prabhakaran, M. Telling, K. Habicht, P. Smeibidl, and K. Kiefer,
Science {\bf 327}, 177-180 (2010).

\bibitem{Hao09} 
%Fermionic Spin Excitations in Two- and Three-Dimensional Antiferromagnets. 
Z. Hao and O. Tchernyshyov,
\prl {\bf  103}, 187203 (2009).

\bibitem{Balents10}
L. Balents, Nature (London), {\bf 464}, 199 (2010)

\bibitem{deVries09} 
%Scale-Free Antiferromagnetic Fluctuations in the s=1/2 Kagome Antiferromagnet Herbertsmithite
M. A. de Vries, J. R. Stewart, P. P. Deen, J. O. Piatek, G. J. Nilsen, H. M. R\o{}nnow, and A. Harrison, \prl {\bf 103}, 237201 (2009).

\bibitem{Fak12} 
%Kapellasite: A Kagome Quantum Spin Liquid with Competing Interactions
B. F\aa{}k, E. Kermarrec, L. Messio, B. Bernu, C. Lhuillier, F. Bert, P. Mendels, B. Koteswararao, F. Bouquet, J. Ollivier, A. D. Hillier, A. Amato, R. H. Colman, and A. S. Wills, \prl {\bf 109}, 037208 (2012). 

\bibitem{Coldea01}
%Experimental Realization of a 2D Fractional Quantum Spin Liquid
R. Coldea, D. A. Tennant, A. M. Tsvelik, and Z. Tylczynski, \prl {\bf 86}, 1335 (2011).

\bibitem{Han12}
%Fractionalized excitations in the spin-liquid state of a kagome-lattice antiferromagnet
T.-H. Han, J. S. Helton, S. Chu, D. G. Nocera, J. A. Rodriguez-Rivera, C. Broholm, and Y. S. Lee, Nature (London) {\bf 492}, 406-410 (2012). 

\bibitem{Freedman10}
%Site Specific X-ray Anomalous Dispersion of the Geometrically Frustrated Kagomé Magnet, Herbertsmithite, ZnCu3(OH)6Cl2 
D. E. Freedman, T. H. Han, A. Prodi, P. Mueller, Q.-Z. Huang, Y.-S. Chen, S. M. Webb, Y. S. Lee, T. M. McQueen, and D. G. Nocera, J. Am. Chem. Soc. {\bf 132}, 16185-90 (2010).

\bibitem{Baker12}
%Spin dynamics of molecular nanomagnets unravelled at atomic scale by four-dimensional inelastic neutron scattering
M. L. Baker, T. Guidi, S. Carretta, J. Ollivier, H. Mutka, H. U. Güdel, G. A. Timco, E. J. L. McInnes, G. Amoretti, R. E. P. Winpenny, and P. Santini, \natphys {\bf 8}, 906-911 (2012).

\bibitem{Tamura09}
%Variety of valence bond states formed of frustrated spins on triangular lattices based on a two-level system Pd(dmit)2
M. Tamura and R. Kato, Sci. Technol. Adv. Mater. {\bf 10} 024304 (2009). 

\bibitem{Pratt11}
%Magnetic and non-magnetic phases of a quantum spin liquid
F. L. Pratt, P. J. Baker, S. J. Blundell, T. Lancaster, S. Ohira-Kawamura, C. Baines, Y. Shimizu, K. Kanoda, I. Watanabe, and G. Saito, Nature (London) {\bf 471}, 612 (2011).

\bibitem{Blundell04}
%Organic and molecular magnets
S. J. Blundell, and F. L. Pratt, \jpcm {\bf 16}, R771-R828 (2004).

\bibitem{Sheckelton12}
%Possible valence-bond condensation in the frustrated cluster magnet LiZn2Mo3O8
J. P. Sheckelton, J. R. Neilson,	 D. G. Soltan, and T. M. McQueen, \natmat {\bf 11}, 493–496 (2012).

\bibitem{Masuda08}
%Magnetic Excitation in Artificially Designed Oxygen Molecule Magnet
T. Masuda, S. Takamizawa, K. Hirota, M. Ohba, and S. Kitagawa, \jpsj {\bf 77}, 083703 (2008).

\bibitem{Canevet10}
%Strong interplay between magnetic and structural properties in the spin-1/2 chain molecular compound D-F5PNN
E. Can\'evet, B. Grenier, Y. Yoshida, N. Sakai, L.-P. Regnault, T. Goto, Y. Fujii, and T. Kawae, \prb {\bf 82}, 132404 (2010).

\bibitem{Yamaguchi13}
%Two-Dimensional Honeycomb Lattice Consisting of a New Organic Radical 2-Cl-6-F-V
H. Yamaguchi, A. Toho, K. Iwase, T. Ono, T. Kawakami, T. Shimokawa, A. Matsuo and Y. Hosokoshi, \jpsj {\bf 82}, 043713 (2013).

\bibitem{Torardi85}
%Synthesis, crystal structures, and properties of lithium zinc molybdenum oxide (LiZn2Mo3O8), zinc molybdenum oxide (Zn3Mo3O8), and scandium zinc molybdenum oxide (ScZnMo3O8), reduced derivatives containing the Mo3O13 cluster unit.
C. C. Torardi, and R. E. McCarley, Inorg. Chem. {\bf 24}, 476–481 (1985).

\bibitem{Sheckelton13}
%Local magnetism and spin correlations in the geometrically frustrated cluster magnet LiZn$_2$Mo$_3$O$_8$
J. P. Sheckelton, F. R. Foronda, LiDong Pan, C. Moir, R. D. McDonald, T. Lancaster, P. J. Baker, N. P. Armitage, T. Imai, S. J. Blundell, and T. M. McQueen, {\ttfamily arXiv:1312.0955} (2013).

\bibitem{Suppl} 
See Supplemental Material for details of the Curie-Weiss analysis, valence-bond fits and $E_i=154$~meV neutron scattering spectra.

\bibitem{Abe09}
%Structuralrefinementof T2Mo3O8 (T=Mg,Co,Zn and Mn)and anomalous valence of trinuclear molybdenum clusters in Mn2Mo3O8
H. Abe, A. Sato, N. Tsujii, T. Furubayashi, and M. Shimoda, \jssc {\bf 183}, 379-384 (2010).

\bibitem{Rodriguez08}
%"MACS-a new high intensity cold neutron spectrometer at NIST"
J. A. Rodriguez \etal, % D. M. Adler, P. C. Brand, C. Broholm, J. C. Cook, C. Brocker, R. Hammond, Z. Huang, P. Hundertmark, J. W. Lynn, N. C. Maliszewskyj, J. Moyer, J. Orndorff, D. Pierce, T. D. Pike, G. Scharfstein, S. A. Smee and R. Vilaseca, 
Meas. Sci. Technol. {\bf 19}, 034023 (2008).

\bibitem{Abernathy12}
%"Design and operation of the wide angular-range chopper spectrometer ARCS at the Spallation Neutron Source"
D. L. Abernathy \etal,
%M. B. Stone, M. J. Loguillo, M. S. Lucas, O. Delaire, X. Tang, J. Y. Y. Lin and B. Fultz, 
Review of Scientific Instruments {\bf 83}, 15114 (2012).

\bibitem{Xu13}
%Absolute cross-section normalization of magnetic neutron scattering data
G. Xu, Z. Xu, and J. M. Tranquada,
{\tt arXiv:1305.5521} (2013).

\bibitem{Lovesey84}
S. W. Lovesey, Theory of Neutron Scattering from Condensed Matter (Claredon Press, Oxford, 1984).

\bibitem{Huse88}
%Simple Variational Wave Functions for Two-Dimensional Heisenberg Spin-½ Antiferromagnets
D.~A. Huse and V. Elser, \prl {\bf 60}, 2531-2534 (1988).

\bibitem{Jolicoeur89}
% Spin-wave results for the triangular Heisenberg antiferromagnet
Th.~Jolicoeur and J. C. Le Guillou, \prb {\bf 40}, 2727-2729 (1989).

\bibitem{Flint13}
%Emergent honeycomb lattice in LiZn2Mo3O8
R. Flint and P. A. Lee, \prl {\bf 111}, 217201 (2013).

\bibitem{Messio11}
% Lattice symmetries and regular magnetic orders in classical frustrated antiferromagnets
% Time-reversal-symmetry-breaking chiral spin liquids: a projective symmetry group approach of bosonic mean-field theories
L. Messio, C. Lhuillier, and G. Misguich, \prb {\bf 83}, 184401 (2011); \prb {\bf 87}, 125127 (2013).

\end{thebibliography}
\end{document}